\documentclass[aps,pre]{revtex4}
\usepackage{amsmath,amssymb,bm}
\usepackage{graphicx}
\usepackage{color}
\include{epsf}

\def\(({\left(}
\def\)){\right)}
\def\[[{\left[}
\def\]]{\right]}

\newcommand{\be}{\begin{equation}}
\newcommand{\ee}{\end{equation}}
\newcommand{\bea}{\begin{eqnarray}}
\newcommand{\eea}{\end{eqnarray}}

\def\LI{1}
\def\BP{5}
\def\FR{6}
\def\ZZ{2}
\def\TU{9}
\def\PR{8}

\begin{document}
\title{Inference in particle tracking experiments by passing messages between images} 

\author{M. Chertkov $^{1,2}$,
  L. Kroc $^1$, F. Krzakala $^{1,3}$, M. Vergassola $^{4}$, and
  L. Zdeborov\'a $^{1}$}

\affiliation{
$^1$ Theoretical Division and Center for Nonlinear Studies, Los Alamos National Laboratory, NM 87545 USA
$^2$ New Mexico Consortium, Los Alamos, NM 87544
$^3$ ESPCI ParisTech and UMR 7083 ``Gulliver'', France
$^4$ Institut Pasteur; CNRS URA 2171, F-75015 Paris, France}

\begin{abstract}
  Methods to extract information from the tracking of mobile
  objects/particles have broad interest in biological and physical
  sciences. Techniques based on simple criteria of proximity in
  time-consecutive snapshots are useful to identify the trajectories
  of the particles. However, they become problematic as the motility
  and/or the density of the particles increases due to
  uncertainties on the trajectories that particles followed
  during the images' acquisition time.  Here, we report
  an efficient method for learning parameters of the dynamics of the
  particles from their positions in time-consecutive images. Our
  algorithm belongs to the class of message-passing algorithms,
  known in computer science, information theory and statistical
  physics as Belief Propagation (BP). The algorithm is
  distributed, thus allowing parallel implementation suitable for
  computations on multiple machines without significant inter-machine
  overhead. We test our method on the model example of particle
  tracking in turbulent flows, which is particularly challenging due
  to the strong transport that those flows produce. Our numerical
  experiments show that the BP algorithm compares in quality with
  exact Markov Chain Monte-Carlo algorithms, yet BP is far superior in
  speed. We also suggest and analyze a random-distance model that
  provides theoretical justification for BP accuracy. Methods developed
  here systematically formulate the problem of particle tracking and
  provide fast and reliable tools for its extensive range of
  applications.
\end{abstract}

\maketitle
%


Tracking of mobile objects is widespread in the natural
sciences, with numerous applications both for living and inert
``particles''. Trajectories of the particles are to be obtained from
successive images, acquired sequentially in time at a suitable
rate. Examples of living ``particles'' include
birds in flocks \cite{Betal_PNAS_08} and motile cells
\cite{Development_08}. Among inert objects,
nanoparticles \cite{Saxton_08} and particles advected by turbulent
fluid flow \cite{Bod01,Pinton01,Adrian_05} provide two important
examples.  The general goal of tracking particles is to extract
clues about their dynamics and to make inferences about the
laws of motion and/or unknown modeling parameters.

Ideal cases for tracking are those where the density and the mobility
of particles is low and the acquisition rate of images is high.  The
non-dimensional parameter governing the stiffness of the
problem is the ratio $\Lambda=\ell\rho^{1/d}$ of the typical distance
$\ell$ traveled by the particles during the time between
images and the average inter-particle distance $1/\rho^{1/d}$. Here,
$\rho$ is the number density of particles and $d$ is the space
dimensionality. Tracking is rather
straightforward if $\Lambda$ is small: the positions of each particle in two
successive images will be relatively
far from those of all other particles. Trajectories are thus
defined without ambiguity. Such a situation is encountered for instances
of the tracking of nanoparticles \cite{Masson09}.  More generally, 
effective methods are available to
identify the assignment (defined as a one-to-one mapping of the
particles from one image to the next one, i.e., the
set of trajectories for all tracked particles) that is the most
probable \cite{Kuhn55,auction,BSS08}.

\begin{figure}[th]
  \hspace{-1.98cm}
    \resizebox{0.8\linewidth}{!}{\includegraphics{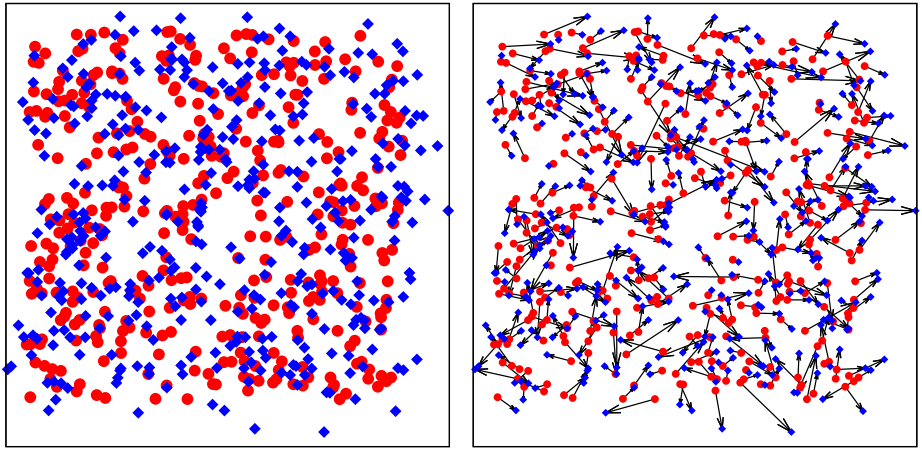}}
    \caption{ A concrete example of particle tracking, with $N=400$
      particles moving from their original positions (red circles) to new ones
      (blue diamonds). In the left figure, two consecutive images
      are superimposed to facilitate comparison of the
      successive positions. Particles are transported by a
      turbulent fluid flow with local stretching,
      shear, vorticity and diffusivity parameters $a^*=0.28$, $b^*=0.54$,
      $c^*=0.24$ and $\kappa^*=1.05$ (see eq.~(\ref{turb})). We focus on
      turbulent transport because of the challenges it poses, yet the methods
      we develop are quite general. The right figure shows the actual
      motion of each particle. Evidently, the simple criterion of particle
      proximity fails to pick the actual trajectories, and the mapping
      of the particles between the two images is intrinsically uncertain.
      Nevertheless, the inference algorithms described here rapidly yield
      excellent predictions ($a=0.32$, $b=0.55$, $c=0.19$ and $\kappa=1.00$)
      for the parameters of the flow.}
\label{Fig0}
\end{figure}

The level of difficulty soars as $\Lambda$ increases: many
sets of trajectories, i.e., many mappings among particles in successive
images, have comparable likelihoods, see Fig.~\ref{Fig0}. The dynamics of the particles ought to be described by
an explicit model, which generally features unknown parameters.
The model defines a probability distribution over the
space of all possible assignments. Contrary to the small $\Lambda$
case, the probability distribution is not necessarily dominated
by a unique assignment. Notwithstanding this uncertainty on the
trajectories, it is expected that useful
information might still be extracted if the number of
tracked particles is sufficiently large. The difficulty is
that all possible assignments must be considered\,: restricting to
the most probable assignment (MPA) generally leads to biased inferences
(see the sequel). Reliable inference requires summing over all possible
assignments with their appropriate weights. Problems with large $\Lambda$
occur in practice; e.g., even with state-of-the-art cameras, particles in
turbulent flow \cite{EB_06} and birds in flocks
\cite{Betal_PNAS_08} often feature ambiguities in the
reconstruction of particles' trajectories. Developing systematic methods to
tackle these cases constitutes our scope here.

The plan of the paper is as follows.
First, we formulate the problem of particle tracking in
terms of a graphical model. We then show that an exact and rapid algorithm for
summing over all possible assignments is unlikely to become available,
as such an algorithm could equivalently
compute the permanent of a non-negative matrix, a problem
that is well-known to be $\#P$-complete~\cite{V:perm_comp}. An approximate
message-passing Belief Propagation (BP) algorithm
\cite{Pearl,MacKay,YFW05} is then introduced, employed and tested.
We also introduce a simplified model
where analytical results and a quantitative sense of the BP
approximations are obtained. The Results section presents
numerical simulations comparing results of BP and
the inference based on the MPA. For the sake of
concreteness, we consider the case of particles passively transported
by a turbulent flow, but the methods are quite general and can be
applied to other situations as well.

\section{Models}
\noindent {\bf Tracking of particles as a graphical model.} Graphical
models provide a framework for inference and learning problems
widespread in machine learning, bioinformatics,
statistical physics, combinatorial optimization and error-correction
\cite{Pearl,MacKay,MM}. The assignment problem involved in the
tracking of particles is conveniently recast as
a weighted complete bipartite graph (see Fig.~2). Nodes are
associated with the $N$ particles in each of two successive images, their
positions being denoted ${\bm x}_i$ and ${\bm y}^j$,
respectively. These $2N$ position vectors ($i,j=1,\ldots , N$)
constitute the experimental data provided. We suppose that a
model for the dynamics of the particles
is available and features a set of unknown
parameters ${\bm \theta}$. Edges between nodes of the bipartite graph
are weighted according to the likelihood that, according to this
model, a particle moves from the initial position ${\bm x}_i$
to the final position ${\bm y}^j$. Specifically, the formula for the
likelihood of an assignment among (non-interacting) particles in two images reads as follows\,:
\begin{eqnarray}
  {\cal L}(\{\sigma\},{\bm \theta})=C\left(\{\sigma\}\right)\,
  \prod_{(i,j)}\left[P_i^j\left(
    {\bm x}_i,{\bm y}^j|{\bm \theta}\right)\right]^{\sigma_i^j}\,.
 \label{likelihood}
\end{eqnarray}
The Boolean variable $\sigma_i^j$ indicates whether the particles $i$
and $j$ are matched
($\sigma_i^j=1$) or not ($\sigma_i^j=0$). The set of the $N^2$
variables $\sigma_i^j$ is denoted by $\{\sigma\}$.  The constraint function
$C\left(\{\sigma\}\right)\equiv\prod_j\delta\left(\sum_i\sigma_i^j,1\right)
\prod_i\delta\left(\sum_j\sigma_i^j,1\right)$, involving Kronecker
$\delta$ functions, enforces the conditions for a perfect matching,
i.e., a one-to-one correspondence between the particles in the two
images. (Situations where the number of particles in the two images can
differ and/or positions of the particles are uncertain are accommodated within the same formalism presented in the
sequel; see Supplementary Information (SI).)
The quantity $P_i^j$ is the transition probability
that a particle at position ${\bm x}_i$ travels to ${\bm y}_j$ in the
time $\Delta$ between images.  The transition probabilities carry all
of the information about the model for the dynamics of the particles.

\smallskip
The tracking inference problem that we address here is to provide fast
and reliable estimates of the model's unknown parameters ${\bm
  \theta}$, which enter the likelihood of the trajectories
via eq.~(\ref{likelihood}).

\begin{figure}[ht]
    \begin{center}
    \resizebox{0.7\linewidth}{!}{\includegraphics{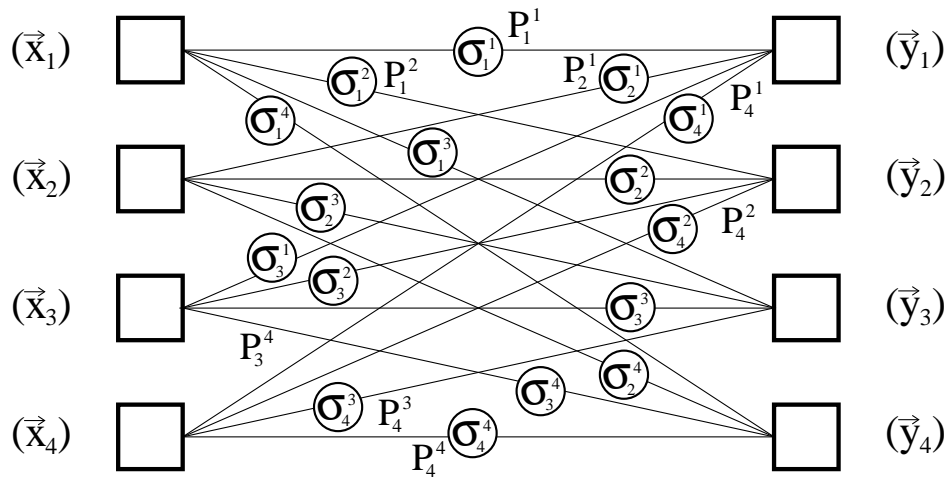}}
    \end{center}
    \caption{The complete bipartite graph for
      the tracking problem. Nodes (squares) denote
      particles in two consecutive images. Edges carry weights
      $P^j_i$ (the likelihood that a particle travels from
      the initial position ${\bm x}_i$ to the final position ${\bm
        y}^j$) and Boolean variables $\sigma_i^j$ (indicating
      whether the nodes $i$ and $j$ in the two images correspond to
      each other ($\sigma=1$) or not ($\sigma=0$)).  Conflicts arise 
    from the constraint that a valid
      assignment is a one-to-one mapping of the particles in the two
      images (expressed by the constraint $C(\{\sigma\})$ in
      eq.~(\ref{likelihood})).}
\label{Fig1}
\end{figure}

The simplest method to infer the unknown parameters ${\bm \theta}$ is
first to identify for any ${\bm\theta}$ 
the MPA, i.e., a
configuration $\{\sigma\}$ satisfying the constraint
$C(\{\sigma\}) = 1$ and having the highest likelihood~(\ref{likelihood}), and
then to maximize the resulting likelihood with respect to ${\bm
  \theta}$. Exact polynomial algorithms~\cite{Kuhn55,auction} solve
the first MPA task. The surprising remark recently made
in~\cite{BSS08,Che08} is that the auction exact algorithm can be reformulated
as a message-passing Belief Propagation (BP) scheme. (Surprise stems
from the fact that BP usually differ from the exact solution of a problem if loops
are present in its graph.)

It is however expected---and confirmed by results described
shortly---that the MPA provides reliable inferences
only for small enough values of the stiffness parameter $\Lambda$
defined in the Introduction. As $\Lambda$ increases,
assignment-dependent entropic factors become important, and MPA inferences
deviate from the actual values of the parameters
${\bm \theta}$. This deviation is understandable because the Bayesian
probability distribution for the parameters, assuming uniform prior
probability for the assignments, involves the full
likelihood~(\ref{likelihood}) marginalized over all possible
assignments\,:
\begin{equation}
P({\bm \theta}|{\bm x_i},{\bm y^j}) \propto \sum_{\{\sigma\}}{\cal L}
\left(\{\sigma\}|{\bm \theta}\right) \equiv  Z\left({\bm \theta}\right) \,.
\label{Z}
\end{equation}
As the stiffness parameter $\Lambda$ increases, many assignments become likely
and the sum~(\ref{Z}) is not dominated by the MPA.

\smallskip
\noindent {\bf Summing over all possible trajectories.} In the vast
majority of cases, no exact algorithm is available to sum over all
possible states of a graphical model. Equation~(\ref{Z}) is no
exception, as it can also be seen as a sum over all possible
permutations of the lower indices $i$ into the upper indices $j$. It
is then recognized that computing the likelihood $Z$ (also known as
partition function in statistical physics) is equivalent to computing
the permanent of the matrix $P_i^j$, a well-known $\#P$-complete
problem~\cite{V:perm_comp}. However, the matrix $P_i^j$ in our
sum~(\ref{Z}) is non-negative and the permanent of non-negative
matrices was the first $\#P$-complete problem discovered to be
solvable by a Fully Polynomial Randomized Approximation Scheme (FPRAS)
\cite{JSV:poly_perm}. The complexity of the
original FPRAS
algorithm is $O(N^{11})$. We significantly accelerated (to $O(N^3)$) and
simplified the original Markov Chain Monte-Carlo (MCMC) algorithm of
\cite{JSV:poly_perm} without observable deterioration of its quality
(see SI). In the Results section, we use this simplified
version to assess the accuracy of our BP approximation while in the SI we
compare performance of BP to the MCMC scheme.

Among the possible approximations to compute the permanent of a
matrix, Belief Propagation (BP) has a special status because of its
aforementioned exactness for the MPA problem
\cite{BSS08}. Moreover the BP algorithm is very fast, scaling as
$O(N^2)$ in its basic form and linearly if, for each particle, only a
limited number of nearby particles is considered. We shall therefore
pursue the development of BP to approximate the sum in (\ref{Z}) and
then assess its validity through numerical simulations
and the simplified ``random distance'' model discussed shortly.

The starting point of the BP approach is the remark that the convex
Kullback-Leibler functional \begin{eqnarray}
{\cal F}\{b(\{\sigma\})\}\equiv \sum_{\{\sigma\}} b(\{\sigma\})
\ln{ \frac{b(\{\sigma\})}{{\cal L}(\{\sigma\})} }\,,
\label{Gibbs} \end{eqnarray} has a unique
minimum at $b(\{\sigma\})={\cal L}(\{\sigma\})/Z$ (under the
normalization condition $\sum_{\{\sigma\}} b(\{\sigma\})=1$), where
${\cal L}$ is defined in eq.~(\ref{likelihood}) and $Z$ in eq.~(\ref{Z}).  The corresponding
value of the functional ${\cal F}$ is the log-likelihood (free
energy), ${\cal F}=-\ln Z$. This remark
constitutes the basis for variational methods (see \cite{MacKay}),
where the minimum of the functional is sought in a restricted class of
functions. BP (and the corresponding approximation for the free
energy, named in \cite{YFW05} after Hans Bethe) involves an ansatz of
the form
\begin{eqnarray}
 b(\{\sigma\})\approx
 \frac{\prod_i b_i({\bm \sigma}_i)\prod_j b^j({\bm \sigma}^j)}{
 \prod_{(i,j)} b_i^j(\sigma_i^j)}\,,
\label{BP_Belief}
\end{eqnarray}
where each vector ${\bm \sigma}_i\equiv\{\sigma_i^j|j=1,\cdots,N\}$
can be any of the $N$ possible vectors
$(0,\cdots,0,1,0,\cdots,0)$ having exactly one nonzero entry,
and ${\bm \sigma}^j$ is defined analogously. This
ansatz is motivated by the fact that the probability distribution on a
graph with a tree structure, i.e., without loops, takes exactly
the form (\ref{BP_Belief}).
The quantities $b_i$, $b^j$, and $b_i^j$ are called ``beliefs''.
In the absence of loops,
$b_i^j$ represents the probability distribution
for the single Boolean variable $\sigma_i^j$,
and $b_i$ represents the joint probability distribution of the
components of ${\bm \sigma}_i$, which appear in a constraint.
Without loops, the beliefs automatically satisfy the conditions
of marginalization, viz., $b_i^j(\sigma_i^j)=\sum_{{\bm
    \sigma}_i\setminus\sigma_i^j}b_i({\bm \sigma}_i)$, where the sum is
performed over all possible values of the vector ${\bm \sigma}_i$
having the specified value of its component $\sigma_i^j$.

In the presence of loops, the expression (\ref{BP_Belief}) is no longer exact,
and the marginalization conditions must be imposed as additional
constraints. 
The important point
demonstrated in \cite{YFW05} is that minimizing the functional
(\ref{Gibbs}) for the class of functions (\ref{BP_Belief}) (under the
marginalization and the normalization conditions) yields the
message-passing formulation of BP, which was heuristically introduced 
by Gallager for decoding of
sparse codes \cite{Gallager}.

Most commonly, the BP equations are formulated as an iterative
message-passing scheme \cite{MacKay,MM}, where the message $\underline h^{i\to
  j}$ (respectively, $\overline h^{i\to j}$) is sent from from particle $i$
in the first (second) image to particle $j$ in the second
(first) image. In the case of the matching problem, the
messages are determined by solving the following equations:
\begin{equation}
 \overline h^{i\to j} = -\frac{1}{\beta}\ln \sum_{k\neq j} P_i^ke^{\beta \underline h^{k\to i}}\,;\,
\underline h^{j\to i} = -\frac{1}{\beta} \ln \sum_{k\neq i} P_k^je^{\beta \overline h^{k\to j}}\,.
        \label{BP_T}
\end{equation}
The ``inverse temperature'' $\beta$ can be set to unity,
but it is usefully retained to show that the limit $\beta\to
\infty$ yields the exact solution for the MPA problem
\cite{BSS08}.

To solve the BP equations (\ref{BP_T}), the messages
$\left(\{\overline h\},\{\underline h\}\right)$ are
randomly initialized and iteratively updated in order to find a fixed point
of the message-passing equations (\ref{BP_T}). The Bethe free energy then reads
\begin{eqnarray}&& -\beta {\cal F}_{BP}\left(\{\overline h\},\{\underline h\},{\bm \theta}\right)= \sum_{(ij)}\ln
  \left(1+P_i^j e^{{\beta \overline h}^{i\to j}
        +{\beta \underline h}^{j\to i}}\right) - \nonumber \\
  && \sum_i \ln \left( \sum_j P_i^je^{\beta \underline h^{j\to
        i}}\right)-\sum_j \ln \left( \sum_i P_i^je^{\beta \overline h^{i\to
        j}}\right) \,.
\label{free_en}
\end{eqnarray}
The Bethe free energy ${\cal F}_{BP}$, evaluated at the fixed point of
the BP equations (\ref{BP_T}), provides an estimate of the exact
log-likelihood $-\ln{Z({\bm \theta})}$ defined through (\ref{Z}).
Because the most likely set of parameters ${\bm \theta}$
minimizes $-\ln{Z({\bm \theta})}$,
we seek parameters that minimize the estimated free energy.
We perform this minimization using Newton's method in combination with message-passing:
after each Newton step, we update the messages
$\left(\{\overline h\},\{\underline h\}\right)$
according to (\ref{BP_T}) and the current set of parameters ${\bm \theta}$.
As this combined update takes $N^2$ steps and the number of combined step which led to convergence is
$N$-independent,
the running time scales as $O(N^2)$. Note that the
running time can be further reduced to $O(N)$
neglecting the contribution of edges with very small probability
$P^j_i$, i.e. diluting the fully connected bipartite graph.

\smallskip
\noindent {\bf A simplified model to understand the BP approximation.}
The BP approximation is exact only if the underlying graph is a tree,
which is not the case for our fully connected bipartite graph in Fig.~2. In the
Results, we empirically assess the validity of the BP
approximation through numerical simulations. To supplement the numerical
evidence, we introduce
here a simplified ``random distance'' model for which analytical results
can be obtained and used to understand the nature of the BP
approximation.

The ``random distance'' model is defined as follows. First, we
decouple the $N^2$ distances $d_i^j$ between particles $i$ and $j$ by
assuming that they are independent among each other. We then assume
that one permutation $\pi^*$ of the lower indices $i$ into the upper
indices $j$ has a special status, while all other distances $d_i^j$
are drawn independently at random from a given distribution. Namely,
the $N$ distances $d_i^{j=\pi^*_i}$'s are distributed as a Gaussian
(restricted to positive values) with variance $\kappa^*=O(1)$. For each of the
other $N(N-1)$ pairs $(i,j)$, the distances $d_i^j$ are independent
random variables drawn uniformly in the interval $(0,N)$. Units of
length are chosen so that the typical interparticle distance is set to
unity.  Note that any distribution of $d^j_i$ for
$j\neq \pi^*_i$ with a vanishing
derivative at the origin would give the same solution. Indeed, the crucial
property is that, for each particle
$i$, the number of distances $d_i^j$ that are
comparable with the diffusion length scale $\sqrt{\kappa^*}$ is $O(1)$.

The interest of a special permutation $\pi^*$ is that we can
inquire about\,: learning $\kappa^*$ if $\pi^*$ is supposed unknown;
the relevance of entropic factors for the partition function; and the
status of the BP approximation.  The same questions arise for
(\ref{Z}) in the original problem. Note that if there were no special
permutation, then we would obtain the random-link model considered in
\cite{MP85,MP86,MMR04}. Our ``random distance'' model can be solved
exactly in the thermodynamic limit using the replica method, as in
\cite{MP85,MP86}, or using the cavity method, as in \cite{MMR04}. The
main result (see SI) is that the BP
expression of the free energy for the ``random distance'' model is
exact in the limit $N\to \infty$, despite of the short loops in the 
graph of the model. The argument
to prove this result goes as follows\,: (a)~the contribution to the
partition function $(\ref{Z})$ from those permutations of the $j$
indices that contain distances larger than $O(1)$ is negligible;
(b)~as each node has only a fraction $O(1)$ of its $N$ distances being
$O(1)$, the underlying graph is effectively sparse; (c)~because sparse
graphs are locally tree-like and correlations in the matching problem
decay very fast on trees, it follows that the BP approximation is
exact in the thermodynamic limit. We have formalized these statements
within the replica and cavity methods (and rigorous local weak
convergence methods are probably also applicable, as for the random
link model \cite{Aldous}).

The asymptotic exactness found in the ``random distance'' model means that
errors made by BP are caused by correlations among the
inter-particle distances. In a smooth flow (see the sequel) 
particles close to each other in the
first image will also be near each other in the second image;
moreover, the four distances among the two positions in the first
image and the two positions in the second will also be small, or more
generally correlated. This effect is more important in lower
dimensions. Indeed, BP inferences turn out to be
better in the three-dimensional (3D) case than in 2D
and rather inaccurate in 1D (with maximum relative error
about 60\%).

Our replica calculations also show that the ``random distance'' model
presents an interesting phase transition at the diffusivity
$\kappa_c\approx 0.174$. For $\kappa^*<\kappa_c$, the MPA $\pi_{MPA}$ is identical to the special one $\pi^*$ with high probability,
whereas for $\kappa^*>\kappa_c$ the overlap (defined via the Hamming
distance) between the most likely assignment $\pi_{MPA}$ and the
special one $\pi^*$ is extensive, i.e., $O(N)$. The comparison with the
finite-dimensional case is discussed in the Results section.

\begin{figure}
  \hspace{-0.7cm}\resizebox{0.9\linewidth}{!}{\includegraphics{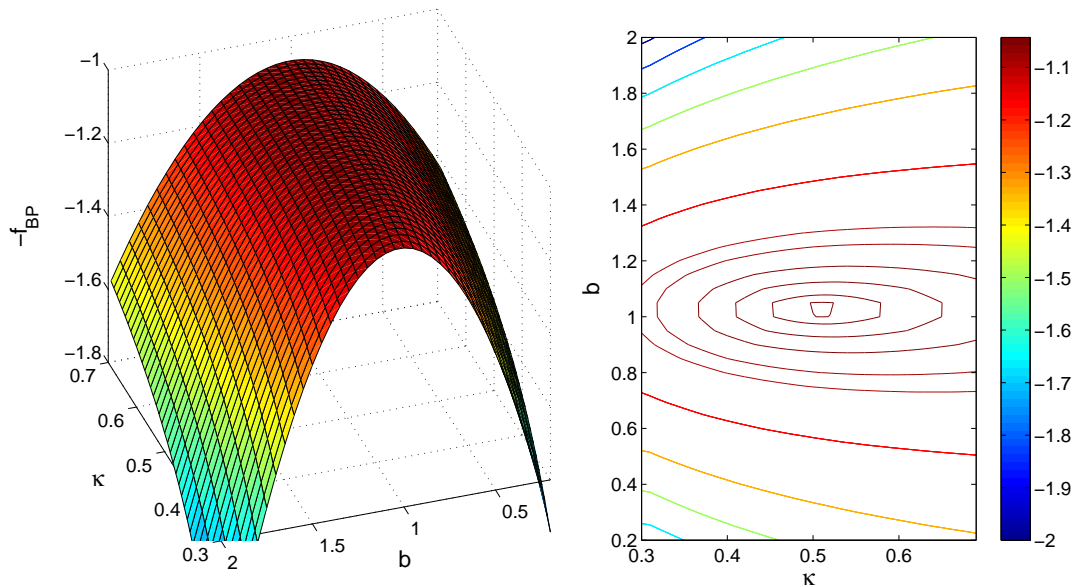}}
  \caption{ A realization of a 2-dimensional flow with
    $a^*=b^*=c^*=1$, $\kappa^*=0.5$ and $N=200$
    particles. Left: The BP Bethe free energy as a function of the
    diffusivity $\kappa$ and the shear $b$, where every point is
    obtained by minimizing with respect to the stretching $a$ and the
    vorticity $c$ of the flow. Right: The same free energy in a contour plot,
    showing the maximum close to $b=1$ and $\kappa=0.5$.  The maximum
    is achieved for $a_{BP}=1.148(1)$ $b_{BP}=1.026(1)$
    $c_{BP}=0.945(1)$, $\kappa_{BP}=0.509(1)$, where the parentheses
    indicates numerical error on the third digit.}
\label{Fig1.1}
\end{figure}

\begin{figure}
  \hspace{-0.7cm}\resizebox{0.9\linewidth}{!}{\includegraphics{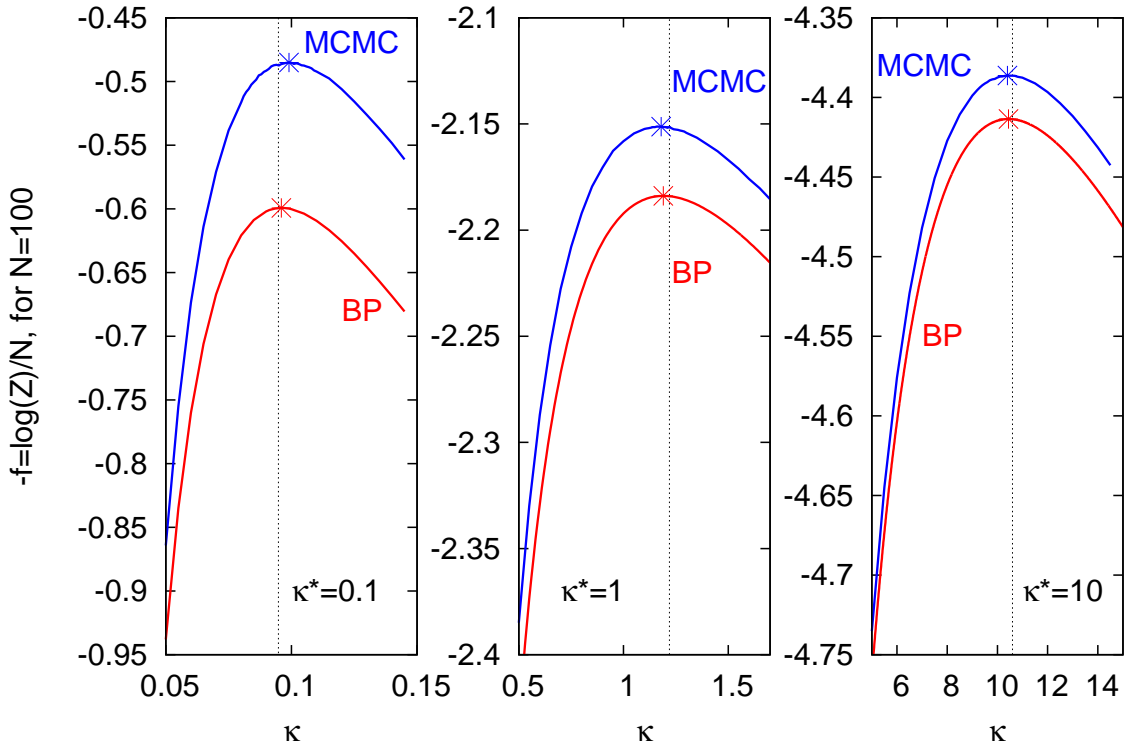}}
  \caption{ The BP estimate of the log-likelihood $-{\cal F}_{BP}$
    and the Monte-Carlo Markov-Chain estimate plotted {\it vs.}\ the diffusivity
    $\kappa$ with $\kappa*=0.1, 1, 10$ for $N=100$ particles diffusing
    in 3D. Although the BP log-likelihood is significantly lower than the
    MCMC one, the estimate of the diffusivity is extremely good.
    The vertical lines mark the diffusivity computed based on the knowledge
    of the actual displacements. The BP algorithm compares favorably in estimation of the maximum with the basically exact Markov Chain Monte-Carlo algorithm, while being far superior in speed 
(for details on speed comparison see SI).}
\label{Fig2}
\end{figure}

\section{Results}

Our analysis has been quite general so far. To concretely assess the
validity of BP, it is now necessary to
specify the model appearing in the likelihood (\ref{likelihood}),
namely the probability $P_i^j$ for the transition from position ${\bm
  x}_i$ to ${\bm y}^j$ in the image acquisition time $\Delta$. We
decided to focus on the tracking of particles in turbulent flow for
three reasons. First, the problem is highly relevant as
an important part of modern experiments in fluid dynamics is based on
the tracking of particles either in
simple~\cite{Bod01,Pinton01,Adrian_05} or complex~\cite{Ouellette09}
flows.
Second, all the algorithms used so far for reconstructing the flow
from images of multiple particles have ignored the probabilistic
structure of the possible assignments. As discussed in the
review \cite{EB_06}, the general approach is to search for a single
matching. Criteria based on proximity and/or minimal acceleration
identify for each particle its ``best'' mapping in the successive
time-image. Conflicts where two or more particles in the first image
are assigned to the same particle in the second image are resolved by
various heuristics. The simplest option is to give up on those
situations where a conflict arises; the most
elaborate solution is to compute the assignment with the minimal cost
in terms of proximity or acceleration.  The bottomline is that one is
always left with a single assignment, which leads to
predictions that are effective at low density of the particles but
rapidly degrade as their density increases \cite{EB_06}.  Third, the
laws of motion of the particles are well-known.  Indeed, if particles
are sufficiently small and chosen of appropriate (mass) density, their
effect on the flow is negligible and they are transported almost
passively. The (number) density of particles is usually rather high
and a single snapshot contains a large number thereof. The
reason is that the smallest scales of the flow ought to be
resolved. Furthermore, turbulence is quite effective in rapidly
transporting particles so that the acquisition time between
consecutive snapshots should be kept small. Modern cameras have
impressive resolutions, on the order of tens of thousands frames per
second. The flow of information is huge: $\sim {\rm Gigabit}/s$ to
monitor a two-dimensional slice of a $(10cm)^3$ experimental cell with
a pixel size of $0.1mm$ and exposure time of $1ms$. This high rate
makes it impossible to process data on the fly unless efficient
algorithms are developed.

\begin{figure}
  \hspace{-1.0cm}
  \resizebox{0.7\linewidth}{!}{\includegraphics{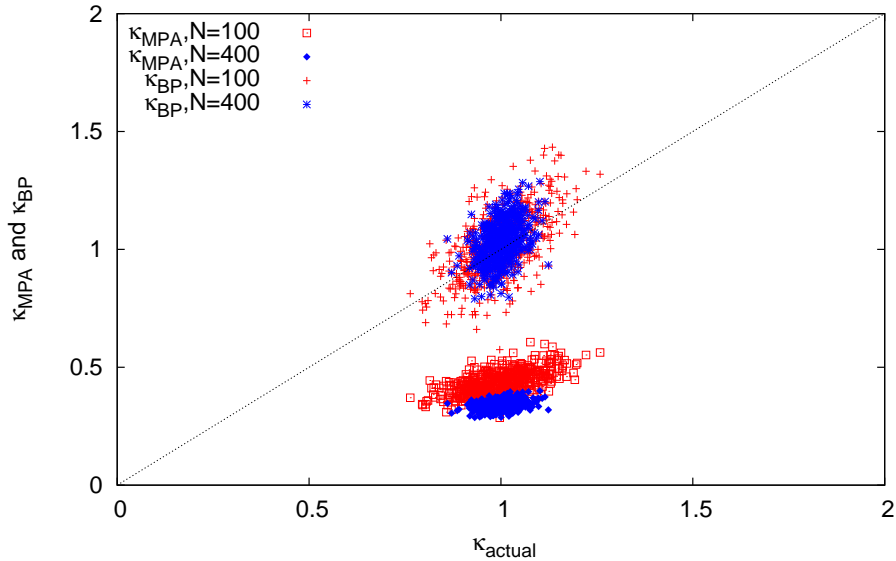}}
  \caption{Scatter plot of the diffusivity estimated
    by BP and by MPA {\it vs.}\ the actual value of the diffusivity.
    Diffusion takes
    place in 3D, with displacements generated using $\kappa^*=1$. The actual
    value of the diffusivity $\kappa_{\rm actual}$ is computed from the actual
    trajectories and is subject to statistical fluctuations.
    The number of tracked particles is $N=100$ (red)
    and $N=400$ (blue) using $1000$ measurements. The BP predictions
    correspond to the maximum of the log-likelihood, as approximated
    by the Bethe free energy (\ref{free_en}) discussed in the
    text. Note the strong underestimation of the MPA estimate, to be
    contrasted with the cloud of BP predictions centered around the
    correct value of $\kappa^*$.}
\label{Fig3.0} 
\end{figure}

\noindent {\bf Tracking of particles in a turbulent flow.} The
likelihood $P_i^j$ in eq.~(\ref{likelihood}) for the transport of
particles in turbulent flow is obtained as follows. Tracked particles
are supposed to be at reciprocal distances smaller than the viscous
scale of the flow. It follows (see \cite{FGV01} for a review) that the
position ${\bm r}_i(t)$ of the $i$-th particle evolves according to
the Lagrangian stochastic equations\,: $\dot{\bm r}_i={\hat s}\cdot
{\bm r}_i+{\bm \xi}_i$. Positions are measured with respect to a
reference point and ${\hat s}$ is the tensor of the velocity
derivatives. In two-dimensional incompressible flow
$a=s_{xx}=-s_{yy}$ is the rate of stretching, $b=(s_{xy}+s_{yx})/2$ is
the shear, and $c=(s_{xy}-s_{yx})/2$ is the vorticity. The stochastic
term ${\bm \xi}_i(t)$ is the zero-mean Gaussian Langevin noise,
describing molecular diffusivity, defined by its correlation function:
$\langle(\xi_i)_{\alpha}(t_1)(\xi_j)_{\beta}(t_2)\rangle=2\kappa
\delta_{ij}\delta_{\alpha \beta}\delta(t_1-t_2)$. The Greek indices
refer to space components. The transition probability
corresponding to the previous transport process is Gaussian:
\begin{eqnarray}
  &P_i^j({\bm x}_i,{\bm y}^j)=  ({\rm det}\, M)^{-\frac{1}{2}} \exp{\left( -\frac{1}{2} {\bm r}^\alpha(M^{-1})^{\alpha\beta} {\bm r}^\beta \right)}\,; \label{eq:prob}\\
  &{\bm r}= {\bm y}^j - W(\Delta)\,  {\bm x}_i\,; \\
  &M= \kappa\,  W(\Delta) \left[ \int_0^\Delta W^{-1}(t) W^{-1,T}(t) \, {\rm d}t \, \right] \,  W^T(\Delta)\, , \label{turb}
\end{eqnarray}
where $W(t)=\exp(t\, \hat s)$
  \footnote{We assumed that the velocity gradients 
  do not change significantly
  between two images. $W(t)$, which is
  generally a time-ordered exponential, is then an ordinary matrix
  exponential. This simplifying assumption can be relaxed, thus
  allowing extension of the
  technique to acquisition times comparable to the viscous scale of turbulence.}
and $\Delta$ denotes the image acquisition time. The general problem of
estimating the unknown parameters ${\bm \theta}$ now takes the 
special form of inferring the
components of the tensor $\hat s$ and the diffusivity $\kappa$.

We tested the BP inference method on numerically simulated data. To 
compare different system sizes, we placed the $N$ particles
at random in a $d$-dimensional box of size $L=N^{1/d}$, i.e., the
average density equals unity. Particles are then displaced
independently following the
probabilistic distribution (\ref{eq:prob}) with a set of parameters
$a^*N^{-1/d}$, $b^*N^{-1/d}$, $c^*N^{-1/d}$, $\kappa^*$. The timescale
was chosen as the acquisition time $\Delta=1$ and the parameters $a$,
$b$, $c$ of the flow were rescaled by $N^{-1/d}$, so that the particle
displacements in the acquisition time are $O(1)$ for all choices of
$N$. Fig.~3 shows the BP free energy
(\ref{free_en}) as a function of the shear $b$ and the diffusivity
$\kappa$ for $N=200$ particles. The curvature around the minimum of
the exact free energy is inversely related to the statistical error in
the estimation of the parameters. Figure~3 clearly shows that
the most problematic parameter is the diffusivity
$\kappa$, as confirmed by all of the numerical simulations
we performed.

In the next three figures, we concentrate on the
purely diffusive regime (where $a^*=b^*=c^*=0$); subsequently, we return to the
general case.  Note that light scattering experiments provide an
established measurement method for biological systems
\cite{Magde,Elson,ElsonII,Rigleretal}. The technique is not commonly
used in fluid dynamics experiments because  the dispersion of the
particles is much faster and the typical illumination level is too weak.

In Fig.~4 we compare the BP results with
results from the fully polynomial randomized approximation scheme
based on the Monte Carlo Markov Chain (MCMC) method. As computed by MCMC
(which is guaranteed to be a FPRAS), the maximum of $Z(\kappa)$
coincides with the true value $\kappa^*$.
From the large deviation theory interpretation of
${\cal F}$, it follows that the statistical error of the estimation of $\kappa$
is $\sigma= 1/[\sqrt{N {\cal
    F}''(\kappa)}]$. Figure~4 confirms the expectation that the
curvature at the minimum of the free energy
decreases with the diffusivity constant $\kappa$.

In Fig.~5 we compare the estimates of the diffusivity
using BP ($\kappa_{BP}$) and using MPA ($\kappa_{MPA}$). The actual
value $\kappa_{\rm actual}$ (respectively, the MPA estimate $\kappa_{MPA}$)
is computed as the mean-square
displacement of the particles on the actual (resp., the most probable)
trajectories of the particles.  The key conclusion to be drawn from
Fig.~5 is that MPA largely
underestimates the diffusivity, whereas the BP method is 
accurate.

Fig.~6 gives a quantitative sense of the BP accuracy as a

function of the diffusivity. The upshot of the curves is that MPA
gives accurate estimates only at extremely low
diffusivities. Conversely, as the diffusivity increases and the
overlap among possible assignments becomes important, the quality of
MPA predictions degrades very rapidly. Results from a similar study
for 2D flows is presented in Fig.~7; vectorial parameters are
again found to be computed efficiently by the BP method.

\begin{figure}[ht]
  \hspace{-1.0cm}
  \resizebox{0.75\linewidth}{!}{\includegraphics{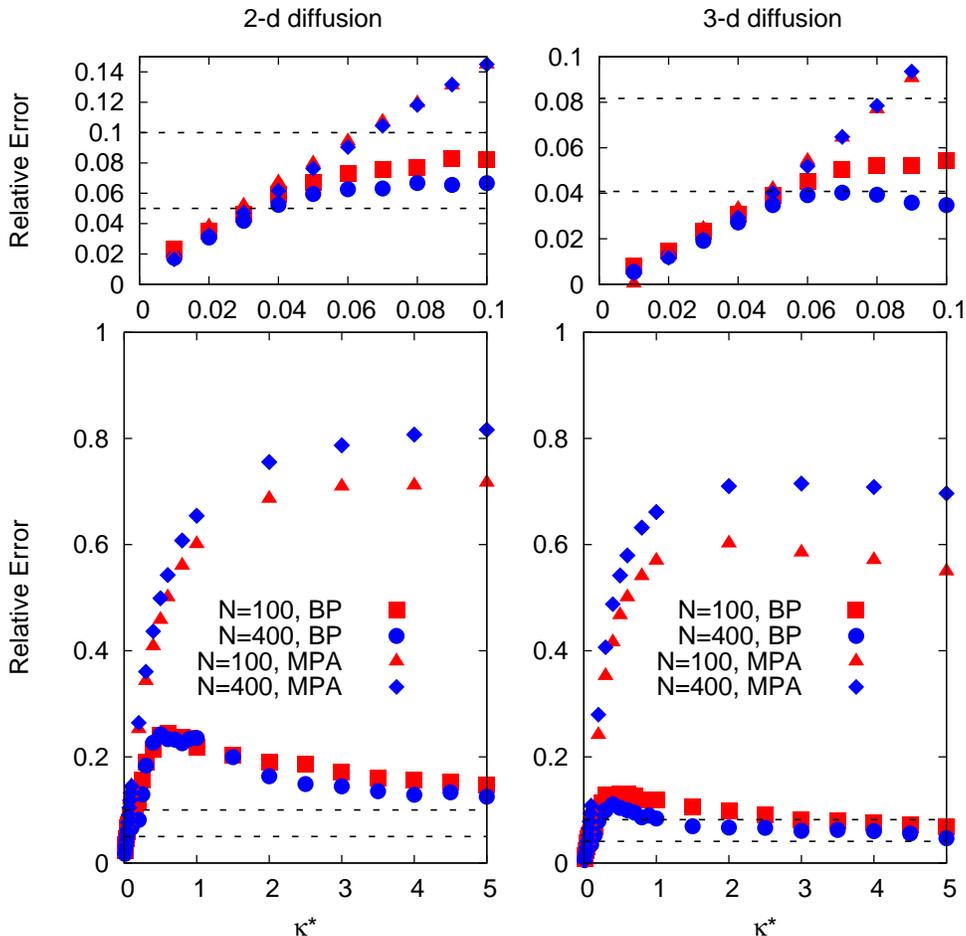}}
  \caption{The relative error $\sqrt{\sum
      (\kappa_{BP/MPA}-\kappa_{\rm actual})^2/K}/\kappa_{\rm actual}$
    in the estimates of the diffusivity over $K$ measurements
    {\it vs.}\ the actual value of the diffusivity $\kappa^*$.
    Circles and squares
    refer to Belief Propagation (BP) while triangles and lozenges
    refer to the Most Probable Assignment (MPA).  $\kappa_{\rm
      actual}$ is the actual value of the mean-square displacement of
    the particles, i.e., it includes fluctuations around $\kappa^*$ due
    to the finite number $N$ of particles. The data are averaged over
    $1000$ (for $N=100$) and $250$ (for $N=400$) realizations and
    compared to relative the statistical error $\sqrt{2/{dN}}$ (dashed
    horizontal lines). The case of two-dimensional (2D) diffusion is
    shown on the left side, and the 3D case is on the right side. The top
    figures are zooms into the low diffusivity region.}
\label{Fig3} 
\end{figure}

\begin{figure}[ht]
  \hspace{-1.0cm}
  \resizebox{0.8\linewidth}{!}{\includegraphics{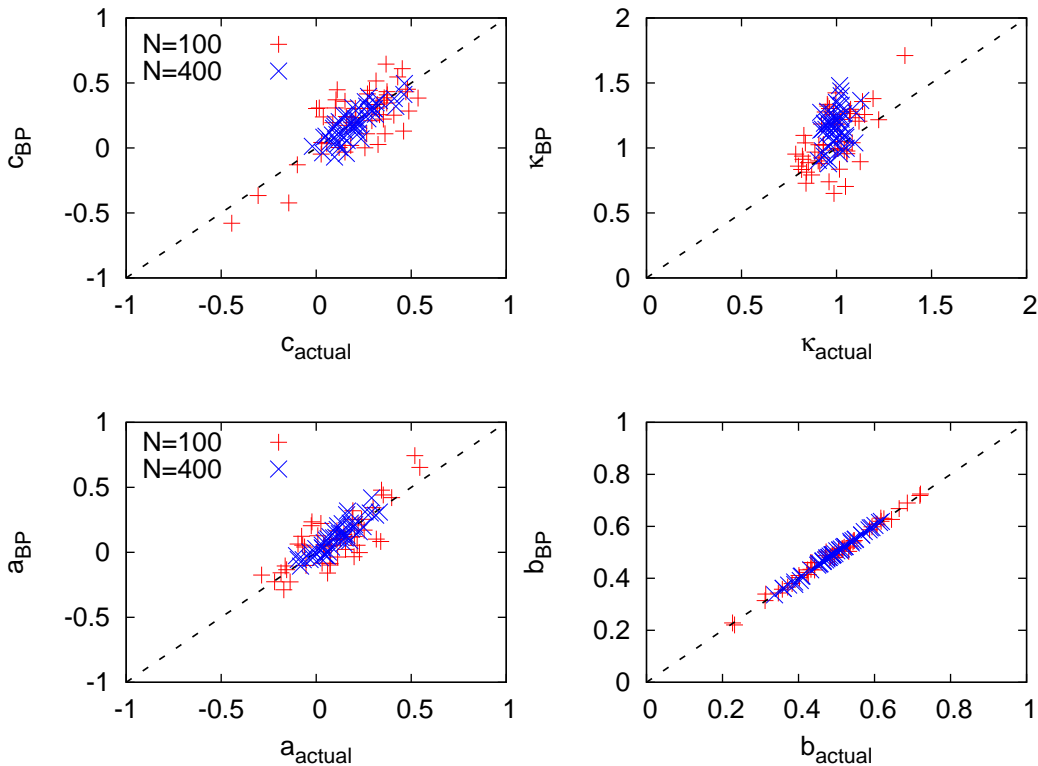}}
  \caption{Scatter plot of the parameter estimations
      using the BP method in the case of a 2D incompressible flow with
      the rate of stretching $a^*=0.1$, the shear $b^*=0.5$, the vorticity
      $c^*=0.2$ and the diffusivity $\kappa^*=1$. Red (blue) points refer
      to the case of a number of tracked particles $N=100$ ($N=400$)
      and the number of measurements is $50$.}
\label{Fig4}
\end{figure}

Finally, Fig.~6 (top panel) indicates that the
phase transition in the exactness of the optimal assignment, which
was previously found for our ``random distance'' model, is smeared out
in the finite dimensional case (or it happens only at
$\kappa<0.01$). This effect is traced to the fact that, in 
finite dimension, there are always several pairs of particles
at distance $o(1)$ that get confused with the diffusion (when
$\kappa^*=O(1)$). It follows that the probability for
$\pi_{MPA}=\pi^*$ is always smaller than unity.

\section{Discussion}

The message-passing algorithms discussed here were shown to ensure
efficient, distributed and accurate learning of the parameters governing the
stochastic map between two consecutive images recording the positions of many
identical particles. The general method was illustrated in a model
relevant for the tracking of particles in fluid dynamics. It was shown that
parameters of the flow transporting the particles could be efficiently and
reliably predicted even in situations where a strong uncertainty in the
particles' trajectories is present.

We introduced and compared two techniques to approximate the
likelihood that the dynamics of the particles is compatible with the
displacements observed in the experimental snapshots. The first is
based on finding the most probable trajectories of the particles
between the times of the two images. The second corresponds to
evaluating the probabilistically weighted sum over all possible
trajectories. The latter is a $\#P$-complete problem and its solution
is approximated by Belief Propagation (BP), as implemented via a
message-passing algorithm. BP was shown to become exact for the
simplified ``random-distance'' model we introduced here. In
general, the effect of loops in the graphical model for the tracking
problem remains nonzero even in the thermodynamic limit of a large number
of tracked particles. Preliminary analysis of the loop corrections to
BP did not display any immediately visible structure, yet detailed
analysis of this point is left for future work. 
Another interesting direction is the development of learning
algorithms (both MCMC and message-passing) specifically designed to
provide estimations of appropriate observables, e.g. the sum of the
square of the distances traveled by the particles, whence the
parameters of the dynamics, e.g. the diffusivity, can be
estimated. This could lead to further reductions in the
computational time and it will be of interest to test whether the
superiority in speed we found here for BP as compared 
to the FPRAS Monte-Carlo scheme (see SI) still holds. 
The price of a single observable is that the
log-likelihood curves in Fig.~4 offer more complete information, namely a 
systematic way to gauge the error bars on the 
inferred parameters.

The algorithms presented here can be carried over to the
tracking of other types of ``particles'', e.g., those of biological
interest. Motile bacteria in colonies, eukaryotic cells or fluorescent
bio-molecules provide relevant examples. As it was stressed here, our
methods are poised to deal with dense conditions where a strong
overlap among the various particles' trajectories are present. An
additional use 
of the techniques
introduced here is to compare different models of transport,
e.g., purely diffusive, directed, active, etc. The Bethe expression
given by the BP approximation can be taken as an
approximation for the log-likelihood of the various models, which are
then compared by standard model selection tests. 
The validity of the various models postulated for the dynamics of
the tracked objects can thus be quantitatively compared.

Our main conclusion is that the BP method gives accurate results and its computational burden is comparable to identifying the most probable trajectories. 
The accuracy of BP was shown to compare extremely well with exact results
and to improve rapidly as the dimensionality of the problem
increases. The BP-based technique allows generalization to reconstruction of a
multi-scale from particle images in two sequential snapshots.  BP can
also be adapted to the case where trajectories of the particles are
reconstructed from their positions in several ($>2$) images (see
SI). In conclusion, the formulation of particle tracking as an
inference problem permits tackling it systematically and introducing
message-passing methods that are highly effective in diverse
applications.

\begin{acknowledgments}

  This material is based upon work supported by the National Science
  Foundation under 0829945 (NMC). The work at LANL was carried out
  under the auspices of the National Nuclear Security Administration
  of the U.S. Department of Energy at Los Alamos National Laboratory
  under Contract No. DE-AC52-06NA25396. Work at IP was partially supported
  by the CNRS program ``Interface''.

\end{acknowledgments}

\newpage

\appendix

\section{Implementation Details}
\label{impl}

\paragraph*{\bf Generating the data}

To generate the set of positions $\bm x_i$ in the original image we
proceed as follows.  We place $N$ particles with positions selected
uniformly at random in a $d$-dimensional box of size $N^{1/d}$, i.e.
each of the $d$ coordinates of any of the $N$ particles is i.i.d. on
the interval $(0,N^{1/d})$.  We choose the set of governing
parameters, ${\bm\theta}=(\kappa,{\hat s})$, describing the
diffusivity and the flow transporting the particles, in such a way
that the displacements of the particles in the acquisition time of the
images (chosen as the unit of time) is of order unity (we remind that
the unit of length is chosen as the typical inter-particle
distance). Namely, since the typical displacement will be of the order
$s r$ and $r=O(N^{1/d})$, we need that each component of the
matrix $\hat{s}$ be $O(N^{-1/d})$.  We therefore performed the
rescaling $s = s^* N^{-1/d}$, with $s^* = O(1)$. To
produce a stochastic map for the $N$ particles from the original image
to their respective positions in the successive image, we generate a
$d$-uple of coupled Gaussian variables $\bm r_i$ with covariance
matrix $M^{-1}$ according to eq.~(\PR-\TU) and then output $\bm
y_i=\bm r_i+W \bm x_i$, where the evolution operator $W$ is defined in
eq.~(\TU).  Note that this setting guarantees that, for sufficiently
large $N$, the number of particles leaving the box is much smaller
than those staying in the box. For the purpose of our validation,
particles which leave the box are treated as if the box did not exist.

\paragraph*{\bf Computing the actual estimate, $\bm \theta_{\rm actual}$}

Given the set of positions $\bm x_i$ and $\bm y^j$ and the actual
permutation~$\pi^*$ (in practice, $\pi^*$ can be taken as the identity
without any loss of generality), we consider the likelihood eq.~(\LI)
to be a function of the set of parameters $\bm \theta$.  The point in
the $\bm\theta$ space where the likelihood achieves its maximum is
called "the actual estimate" and it is denoted as $\bm \theta_{\rm
  actual}$. The estimate $\bm \theta_{\rm actual}$ is found using the
Newton's method. Note that finite size effects make that $\bm
\theta_{\rm actual}$ typically differ from the respective original
value, $\bm \theta^*$, by deviations that are $O(N^{-1/2})$.

\paragraph*{\bf Computing the MPA estimate, $\bm \theta_{\rm MPA}$}

Computation of the Most Probable Assignment (MPA) estimate is split
into two parts. First, for the given set of parameters $\bm \theta$ we
look for the permutation, $\pi_{\rm MPA}$, which maximizes the
likelihood eq.~(\LI); second, we maximize the resulting likelihood
with respect to $\bm \theta$ using the Newton's method. Exact
algorithms are available to achieve the first
task~\cite{Kuhn55,auction,BSS08}, and we utilize in our
simulations the zero-temperature Belief Propagation algorithm of
\cite{BSS08}, as the latter is considerably faster than other
alternatives. In the zero temperature limit, $\beta\to \infty$, the BP
equations (\BP) become \be \overline h^{i\to j} = -\max_{k\neq
  j}(\underline h^{k\to i} + \ln P_i^k ) \,,\, \quad \quad \underline
h^{j\to i} = -\max_{k\neq i}( \overline h^{k\to j}+ \ln
P_k^j)\,.  \label{BP_0} \ee The MPA matching is reconstructed from the
fixed point of eq (\ref{BP_0}) as follows \be \pi_1(i) = -{\rm
  argmax}_{k}(\underline h^{k\to i} + \ln P_i^k ) \,,\, \quad \quad
\pi_2(k) = -{\rm argmax}_{i}( \overline h^{i\to k}+ \ln
P_i^k)\, . \label{BP_0_full} \ee We initialize the messages $\{\overline
h\},\{\underline h\}$ at random, and solve eqs.~(\ref{BP_0})
iteratively till the consistency condition $\pi_2[\pi_1(i)]=i$ for all
$i=1,\dots,N$ is satisfied. Note that this scheme is faster than
enforcing the convergence of the BP equations (\ref{BP_0}) to a fixed
point. {\it A priori} we could reach a configuration of messages where the
condition is satisfied and the corresponding configuration is not the
best matching because the fixed point was not reached yet. However, in
reality we never observed this situation when the number of particles
is sufficiently large.

\paragraph*{\bf Computing the BP estimate, $\bm \theta_{\rm BP}$}

The task here is to find the minimum of the Bethe free energy
eq.~(\FR) over the set $\bm\theta$, where $\{\overline
h\},\{\underline h\}$ solve eq.~(\BP) at $\beta=1$ and given $\bm
\theta$.  Our implementation of this task is as follows.  We
initialize the algorithm with randomly generated $\{\overline
h\},\{\underline h\}$, solve BP equations iteratively and then
minimize the resulting Bethe free energy using the Newton's
method. Each time the Newton's method calls for the value of the free
energy at a certain value of parameters $\bm \theta$, each BP message
is updated on average $m$ times and then the value of the Bethe free
energy is computed. We typically used $m=5$ or $m=10$. In general this
small number of iterations is not sufficient to reach the fixed point
for each value of the parameters, however, we found empirically that
reaching the fixed point at every step is unnecessary as long as the
Newton's method finally converges (which was always the case in our
implementation).

Mixing the BP iterations with the Newton's method updates accelerates
the algorithm significantly.  The running time of this mixed
implementation is quadratic in the number of particles. Moreover, the
following procedure allows to attain a linear scaling without much of a
quality sacrifice: for each particle
$i$, we retain only those $j$'s such that $P_i^j$ exceeds a predefined small constant $C$. As the
size of the system grows, this truncation identifies only $O(1)$
number of possible partners for each particle resulting in the linear
scaling for the algorithm running time.

\section{Cavity solution for the random distance model}
\label{cavity}

We give here a brief exposition to the cavity solution of the ``random
distance'' model defined in the main text.  More details about the
method itself, applied to a different but related model of random
matching, can be found in
\cite{MP85,MP86,MMR04,MartinMezard05}.

In the random distance model the likelihood (\LI) becomes
\be
{\cal L}(\{\sigma\}|\kappa)=C\left(\{\sigma\}\right)\,
\prod_{(i,j)}\left[P_i^j( d_i^j|\kappa)\right]^{\sigma_i^j}
\left[\frac{1}{N}\right]^{1-\sigma_i^j}\,,
\label{ran_like}
\ee
where $P_i^j( d_i^j|\kappa) = \frac{1}{\sqrt{2 \pi \kappa}}
e^{-\frac{1}{2\kappa} (d_i^j)^2} $.  Note that
$\sum_{(i,j)}(1-\sigma_i^j) = N(N-1)$ and thus the term with $1/N$ in
eq.~(\ref{ran_like}) is contributing as a multiplicative constant and
can be ignored. Hence, the belief propagation equations (\BP) hold for
the random distance model, the Bethe free energy is computed according
to eq.~(\FR), and the respective expression for the ground state
energy is \be E = \sum_{(ij)} \max(0, \overline h^{i\to j}+\underline
h^{j\to i}+\ln P_{i}^j) -\sum_{i} \max_{i\in \partial j} ( \underline
h^{j\to i} + \ln P_{i}^j ) -\sum_{i} \max_{j\in \partial i} (
\overline h^{i\to j} + \ln P_{i}^j ) \, .\label{energy_0} \ee

The belief propagation solution applied to an instance of the random
distance problem becomes asymptotically and statistically exact in the
thermodynamic limit $N\to\infty$. This statement is equivalent to
exactness of the so-called replica symmetric solution made in
\cite{MP85,MP86,MMR04,MartinMezard05} for
a different but related random matching model. We tried to detect
instabilities towards replica symmetry breaking in the random distance
model, but failed to find any. This empirical evidence leads us to
conjecture that one may be able to extend the rigorous local weak
convergence method of \cite{Aldous} or independent combinatorial
proof of \cite{LinussonWastlund03} to the random distance model.

\medskip

\paragraph*{\bf Population dynamics solution}
Population dynamics \cite{MezardParisi01} is a general method to solve
the belief propagation equations on a infinite sample without actually
generating or storing the whole sample. The assumption here is that
all the quantities of interest are self-averaging, i.e., their values
are asymptotically equal to respective averages over large sample. For
the matching problem, a trick introduced in
\cite{MMR04,MartinMezard05} exploits the effective sparseness
of the fully connected factor graph. The main observation is that the
best matching has energy proportional to $N$ whereas a random
permutation has energy proportional to $N^2$, and only permutations
with energy proportional to $N$ contribute to the log-partition
function and other quantities of interest. Edges with $O(N)$ weights
almost never contribute to these configurations. The number $k$ of
edges adjacent to a variable whose weights are smaller than a certain
constant $c$ is described by the following Poissonian distribution
with mean $c$: \be R(k)= {N \choose k} \left(\frac{c}{N}\right)^k
\left(1-\frac{c}{N}\right)^{(N-k)} \to_{N\to \infty}
e^{-c}\frac{c^k}{k!}\, .\label{Poiss} \ee Additionally to these $k$
edges we always consider an edge connecting the variable under consideration 
to its
actual image. We thus end up with a bipartite factor graph with the
connectivity degree $1+k$, where $k$ fluctuates according to
eq. (\ref{Poiss}). As $c\to \infty$ we recover the original problem,
but in practice moderate values of $c$ are sufficient to achieve
numerically the asymptotic regime.

Let us now turn to the cavity equations explaining the relations among
the probabilities of the messages $\underline h^{i\to j}$, $\overline h^{i\to
  j}$ at the fixed point of eq. (\BP) on a very large graph. In fact,
there are two kinds of BP messages, these describing transitions
between $i$ and its actual image and also transitions between $i$ and
nodes which do not correspond to $i$'s actual image. Our notation for
the two distributions are $\tilde Q(h)$ and $Q(h)$, respectively. The
resulting cavity equations are \bea
\tilde  Q(h)  &=&     \sum_k   R(k)  \int \prod_{i=1}^k {\rm d}d_i\,  U(d_i)   \int \prod_{i=1}^k   {\rm d}h_i \, Q(h_i) \,  \delta( h- {\cal F}(\{h_i\}))\, , \label{pop_1} \\
Q(h) &=& \sum_k R(k) \int {\rm d}d_0\, P(d_0) \prod_{i=1}^k {\rm d}d_i
\, U(d_i) \int {\rm d}h_0 \, \tilde Q(h_0) \prod_{i=1}^k {\rm d}h_i \,
Q(h_i) \, \delta( h- {\cal F}(\{h_i\},h_0))\, .\label{pop_2} \eea
The quantity ${\cal F}(\{h_i\})$ satisfies eq. (\BP). Let us recall that
$P(d)=\frac{1}{\sqrt{2 \pi \kappa^*}} e^{-\frac{1}{2\kappa^*} d^2}$ is
the distribution of displacements; and $U(d)=1/c$ for $0<d<c$, and
$U(c)=0$ otherwise.

Population dynamics is the method we utilized to solve the cavity
equations (\ref{pop_1}-\ref{pop_2}). In this method both distributions
$Q(h)$ and $\tilde Q(h)$ are represented by a pool of $N_{\rm pop}$
numbers initialized at random.  In one sweep on the population
dynamics we repeat $N_{\rm pop}$ times the following procedure:
\begin{itemize}
\item{Iteration of eq. (\ref{pop_1}): Draw random number $k$ according
    to eq.~(\ref{Poiss}), draw $k$ random numbers $d_i$ uniformly at
    random from interval $(0,c)$, and draw $k$ random numbers from the
    pool representing the distribution $Q(h)$. Then use eq. (\BP) to
    compute a new $h$ and substitute it on a place of a random element
    in the pool representing $\tilde Q(h)$. Note that in the case when
    $k=0$ from the definition of message $h$, the value of $h$ has to
    be large $h\to \infty$. In practice, we take $h=h_{\rm max}$.}

\item{Iteration of eq. (\ref{pop_2}): Draw random number $k$ according
    to eq.~(\ref{Poiss}), draw $k$ random numbers $d_i$ uniformly at
    random from interval $(0,c)$, and draw $k$ random numbers from the
    pool representing the distribution $Q(h)$. Draw one random number
    $d$ from the Gaussian distribution with variance $\kappa^*$, and
    one random number from the pool representing $\tilde Q(h)$. Then
    use eq. (\BP) to compute a new $h$ and use it to replace a random
    element in the pool representing $Q(h)$.}
\end{itemize}

We iterate the procedure till convergence. Knowing the convergent
$Q(h)$ and $\tilde Q(h)$ allows us to evaluate various sample averaged
quantities of interest, such as the density of the free energy (\FR). Note
that to calculate the first term in eq.~(\FR), the term needs to be
split in two parts, correspondent to the actual $(ij)$ pairs (particle
and its image) and "confused" pairs. In practice we compute each of
the terms in eq.~(\FR) $N_{\rm pop}$ times, each correspondent to new
choice of randomness. Then, we do few (around $10$ in practice) sweeps
to update the pools representing $Q(h)$ and $\tilde Q(h)$ and repeat
the measurement to eliminate sampling errors. Similar scheme also
applies to evaluating average of the second derivative of the free
energy.

Let us also mention  for completeness that the system of cavity equations (\ref{pop_1}-\ref{pop_2}) allows an elegant and compact direct representation in the asymptotic $c\to\infty$ limit,  in the spirit of  \cite{MP85,MP86} where this trick was originally proposed for the random matching model. Following \cite{MP85,MP86} we define a generalized Laplace transform as
\be
e^{-G(l)} \equiv \int_{-\infty}^{\infty} {\rm d}h Q(h) e^{-e^{l-\beta h}} \, ,  ~~~~~
e^{-\tilde G(l)} \equiv \int_{-\infty}^{\infty} {\rm d}h \tilde Q(h)
e^{-e^{l-\beta h}} \, .  \ee
Applying this transform to eq. (\ref{pop_1}-\ref{pop_2}) we arrive at the following set of closed equations for $G(l)$, $\tilde G(l)$:
\bea
\tilde G(l) &=& \int_{-\infty}^\infty {\rm d}y \, e^{-G(y)}
\sum_{p=1}^\infty \frac{(-1)^{(p-1)}}{p! (p-1)!}  e^{p(l+y)}
g_p(\beta)\, , \label{orig}\\
  G(l)&=& \tilde G(l)-\log{\left[ 1 - \int_{-\infty}^\infty
    {\rm d}y \, e^{-\tilde G(y)} \sum_{p=1}^{\infty}
    \frac{(-1)^{p-1}}{p!(p-1)!} e^{p(l+y)} \tilde q_p(\beta) \right]}
\, , \label{second}
\eea
where \be g_p(\beta)=\sqrt{\frac{\pi}{2\beta p}}\, , \quad \quad
\tilde q_p(\beta) = \sqrt{\frac{1}{\kappa^*\beta p +1}}\, .  \ee Note,
that eq. (\ref{orig}) with $\tilde G=G$ was originally derived in
\cite{MP85,MP86} using the replica method for the
random matching problem, while the log-correction on the right hand
side of eq.~(\ref{second}) is specific to our problem, representing
bias brought into the problem by the existence of the special
permutation $\pi^*$.

\medskip

\paragraph*{\bf  Results}

Let us briefly discuss the results obtained by the population dynamics
numerical evaluation of the cavity equations
eq. (\ref{pop_1}-\ref{pop_2}).

\begin{figure}[!ht]
    \resizebox{0.45\linewidth}{!}{\includegraphics{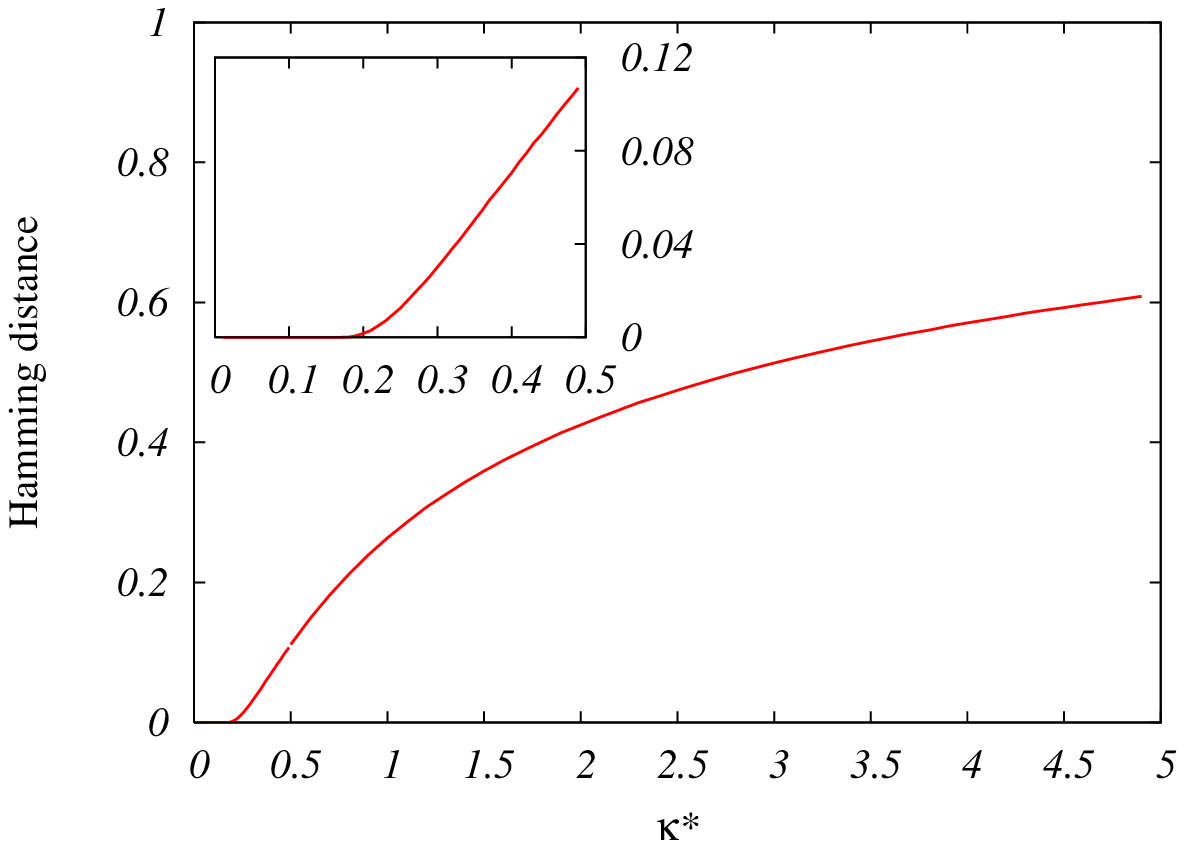}}
    \resizebox{0.45\linewidth}{!}{\includegraphics{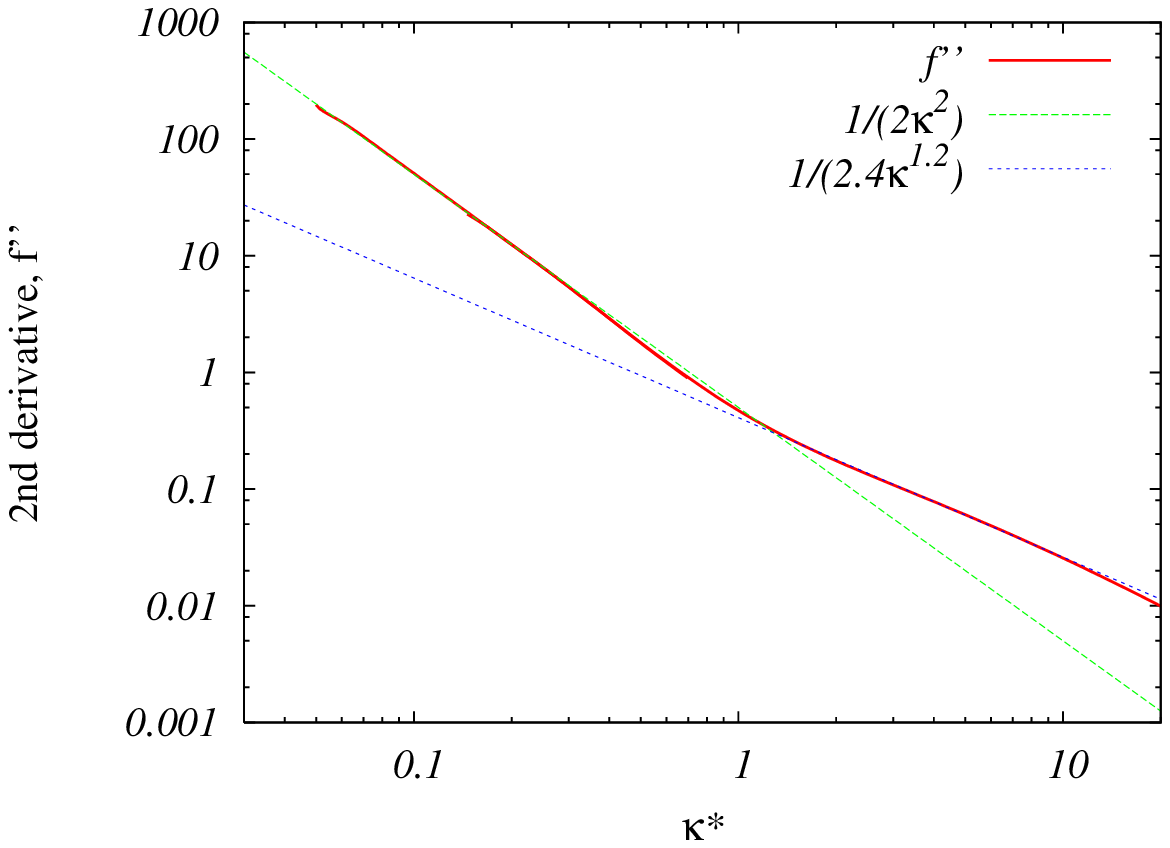}}
    \caption{ \label{Fig:o0T} Left: Result of the zero temperature
      population dynamics with $N_{\rm pop}=10000$, $c=20$, $t_{\rm
        max}=10000$. Distance between the actual and the most probable
      matchings shown as a function of diffusivity $\kappa^*$. Inset:
      Zoom into the region of the phase transition at low
      diffusivity. At about $\kappa_c\approx 0.174(4)$ the most
      probable matching ceases to be equal (asymptotically) to the
      actual matching. Right: The second derivative of the free energy
      at $\kappa^*$ as a function of $\kappa^*$.  We observe the
      change of the slope at $\kappa^*\approx 1$ from
      $1/[2(\kappa^*)^2]$ to a larger value.  }
\end{figure}

First, we compare the most probable matching with the actual matching
and analyze their dependence on the diffusivity $\kappa^*$.  We define
the Hamming distance between the actual and the most probable
matchings as the fraction of edges which are present in the actual
matching but are missed in the most probable matching. The comparison
is shown in Fig.~\ref{Fig:o0T}, where we plot the Hamming distance as
a function of the diffusivity. The inset of the Figure shows a
transition observed at $\kappa^*<0.174(4)$ from the lower diffusivity
phase, where no difference between the most probable and actual
matching were detected, to the high diffusivity phase where the most
probable and actual matchings are distinctly different. The Hamming
distance saturates to $1$ as $\kappa^* \to \infty$. Note that this
phase transition is a property specific to the random distance model
and that it was not observed in our particle simulations in $2d$ and
$3d$ discussed in the main text.

To estimate the most probable value of the diffusion constant $\kappa$
based on the observed data (set of mutual distances) we need to
maximize the full likelihood $Z(\kappa)$, eq.~(\ZZ), with respect to
$\kappa$, i.e. to minimize the free energy of the model.  In the
limit, $N\to \infty$, the minimum of the free energy within the random
distance model is always achieved at $\kappa=\kappa^*$. However, the
minimum gets flatter as the the actual diffusivity $\kappa^*$ grows.
In a finite size system the statistical error in estimating $\kappa^*$
is $1/\sqrt{(N {\cal F}'')}$. Let us remind here that if the actual matching
is known then the statistical error is $\kappa^* \sqrt{(2/N)}$. In
Fig.~\ref{Fig:o0T}, we plot (in log-log-scale) the second
derivative of the free energy as a function of $\kappa^*$. We observe
that for $\kappa^*<1$ the second derivative is roughly
${\cal F}''=1/[2(\kappa^*)^2]$, hence in this regime the statistical error is
of a similar origin as if the actual matching would be known. For
$\kappa^*>1$, the scaling of the second derivative changes to
roughly proportional to $1/\kappa^*$, hence in this regime estimation
of the actual $\kappa^*$ from the full likelihood is more accurate. It
is noteworthy to emphasize that the crossover in the second derivative
(and thus accuracy of the log-likelihood based prediction) is observed
at $\kappa^*\approx 1$, whereas the phase transition in the Hamming
distance (between the actual and most probable matchings) takes place
at the much lower values $\kappa^*=\kappa_c\approx
0.174(4)$.

\section{Comparative analysis of BP and MCMC algorithms}
\label{BPMC}

In the body of the paper we stated the problem of learning the
parameters of the flow/diffusion from particle tracking data in terms
of the partition function of an associated matching problem. One
important point articulated in the manuscript is that, even in the
regime where the most probable matching is very different from the
actual one, computing the partition function enables us to estimate
the parameters accurately. Note, that to the best of our knowledge all
the methods used so far in particle tracking reconstruction/learning,
have been relying solely on  a single matching.

Our results show that BP is both efficient and accurate for
approximating the maximum of the partition function. We have used an
MCMC-learning algorithm, but our original focus was primarily on
assessing the quality of the BP by comparing it to the MCMC
near-to-exact result. To guarantee near-to-exact result from the MCMC
algorithm the running time is impractical. In this Section we give
some additional details and comment also on comparative speed
performance of the two learning algorithms.

MCMC sampling is the most common method for estimating properties of
the Boltzmann distribution. However, standard MCMC approaches sample
from the given distribution aiming to evaluate certain expectation
values and they are not directly suitable for accurate counting
(evaluation of the partition function) required for our purposes. A
considerable speed up in sampling of lower cost configurations is
usually achieved via Metropolis-Hastings implementation of the MCMC,
however, in order to compute the partition function (or the free
energy) one has to numerically integrate the energy function over a
range of temperatures, and this integration is computationally costly.

To resolve this problem we have used a special MCMC algorithm uniquely
designed for this purpose -- a somewhat simplified version of the
Fully Polynomial Randomized Approximation Scheme (FPRAS), originally
introduced for computing permanents (and this is what our partition
function is) in \cite{JSV:poly_perm}. The unabridged version of
this FPRAS algorithm is guaranteed to give a value that is not worse
than $(1+\epsilon)L$ ($L$ being the exact value) in $O(N^{11})$ and
${\rm poly}(1/\epsilon)$ computational time \cite{JSV:poly_perm}. A
description of our simplified algorithm follows.

Let us denote the matrix whose permanent we want to compute by $A$. In
our particular case, $A$ contains the likelihoods of pairs of particles
in different time steps corresponding to each other. The algorithm
works in stages. It starts with a constant matrix $A'$, whose
elements are all equal to the largest value in $A$, $a_{max}$, and
whose permanent is easily computed as $(a_{max})^n n!$.  In each stage
$i$, elements of $A'$ are reduced by a constant factor to
$a'_{uv}=\max\left\{{\rm exp}{(-1/2)}a'_{uv}, a_{uv}\right\}$, thus
bringing $A'$ closer to $A$. Using $S$ (a parameter) samples from all
possible perfect matchings (i.e., permutations), an approximate ratio
$r_i$ is computed of the new value of the permanent of $A'$ and its
old value (before reducing elements of $A'$ in step $i$). The samples
are obtained using a Metropolis-Hastings MCMC, where neighbors of a
state are all matchings with any two pairings swapped, and the
probability of accepting a proposed step is the ratio between products
of values of $A'$ corresponding to new and old matchings. This MCMC is
run for $T$ (a second parameter) steps for each sample. Finally, the
algorithm finishes when the matrix $A'$ becomes sufficiently close to
the original $A$. The approximate value of the permanent is then
obtained as $\prod_i r_i \cdot (a_{max})^n n!$. The two parameters,
$S$ and $T$, determine the time complexity of the algorithm, which is
$O(ST)$.

\begin{figure}[!ht]
    \resizebox{0.45\linewidth}{!}{\includegraphics{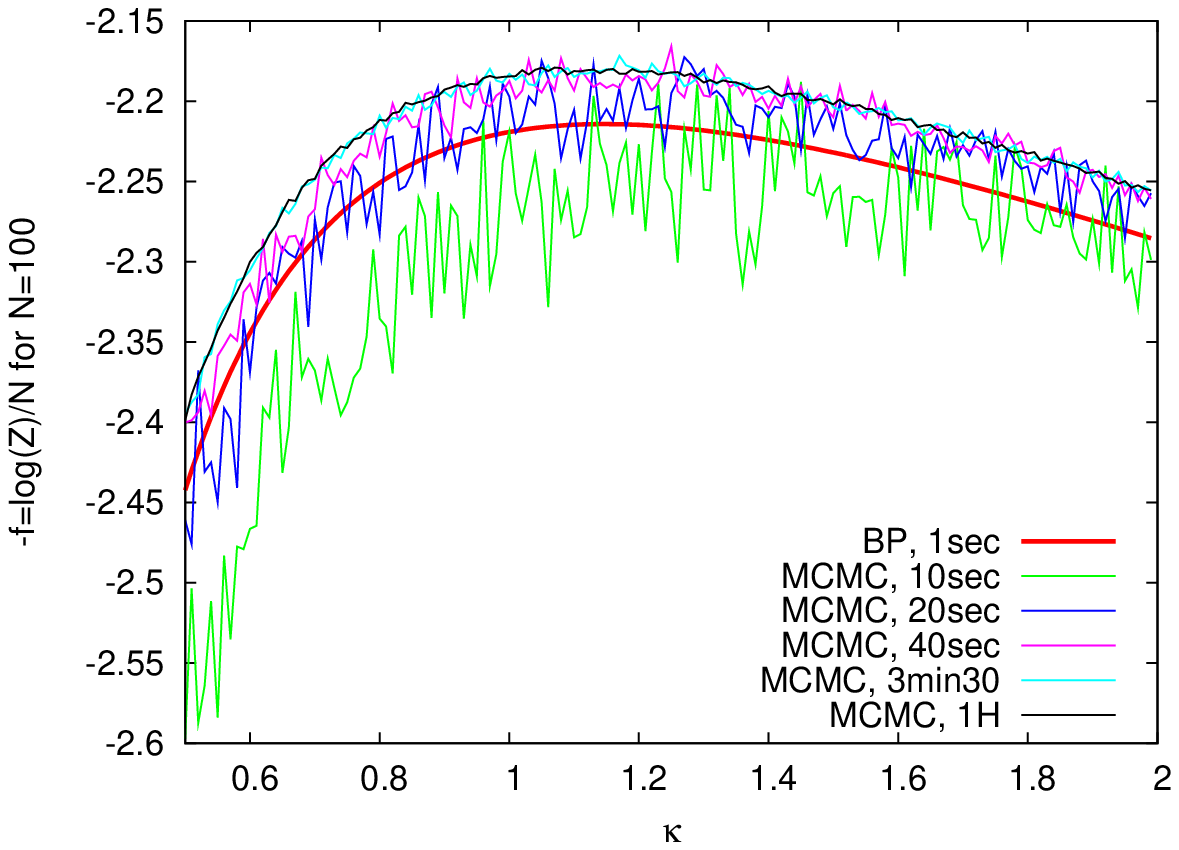}}
    \resizebox{0.45\linewidth}{!}{\includegraphics{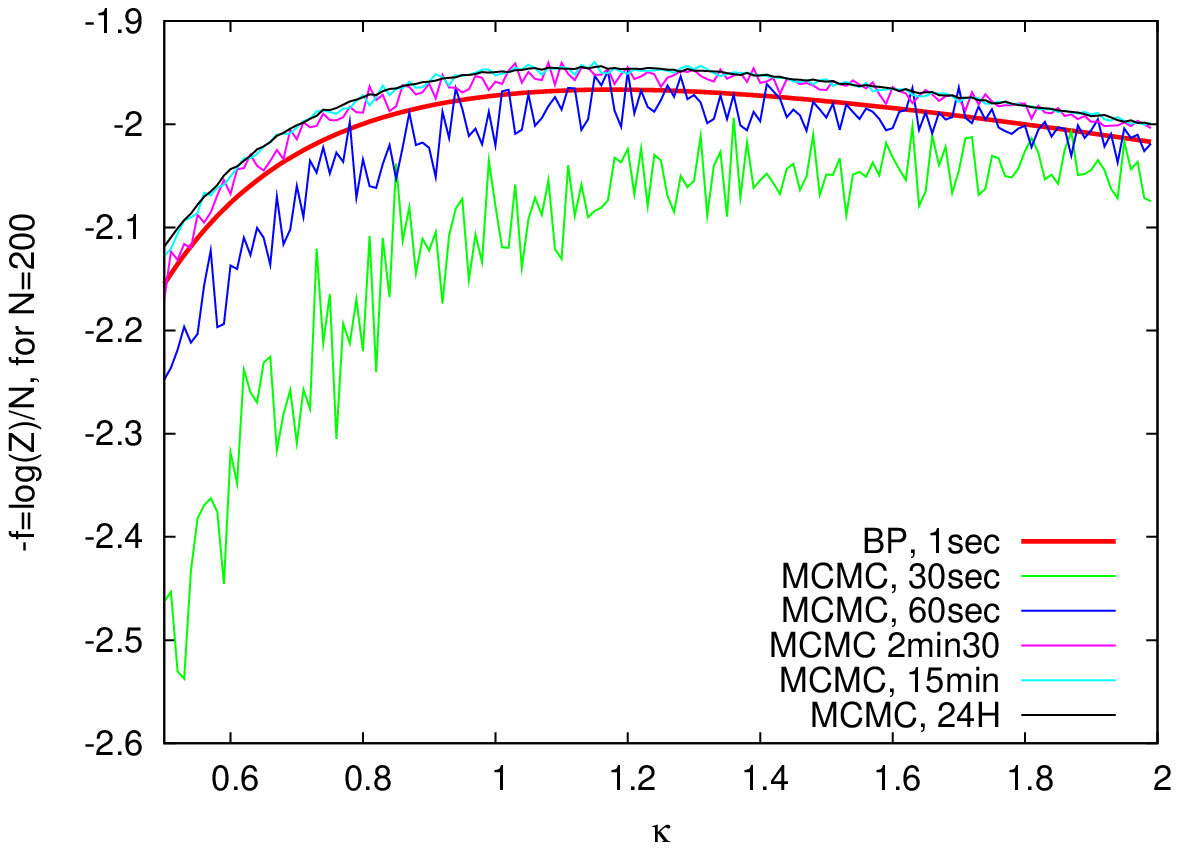}}
    \caption{ \label{Fig:BPMC} Logarithm of the partition function
      computed from the belief propagation algorithm (in red thick
      line) and from the Monte Carlo described in the text. The Monte
      Carlo is guaranteed to find an $\epsilon$ approximation of the
      exact value in $O(N^{11})$ time. We, however, run it for a much
      shorter periods and analyze how does its performance depend on
      the running time. We find that in order to compare with BP in
      the estimation of the maximum, MCMC needs simulations that are
      at least one order of magnitude longer.}
\end{figure}

To facilitate comparison of the MCMC and BP efficiency (speed), we downgraded strict quality requirement of the original FPRAS-MCMC and conducted simulations to access its deterioration in quality with the computational time (and number of samples) decrease.
For this comparison we have used instances of the particle tracking described in Section  \ref{impl}. The results, shown in Fig. \ref{Fig:BPMC}, lead to conclusion that for comparable performance we need about an order of magnitude longer running time for the MCMC. Note, however, that we tested only specific implementation of the MCMC algorithm,
and obviously further comparative analysis of BP-based, MCMC-based and possibly other
learning schemes will be necessary.

We find it useful to complete this Section by 
discussing another important point related to the 
comparison of the MCMC and BP efficiency for evaluating the 
maximum of the partition function. This is a crucial quantity for 
learning as it provides the most likely estimate 
of the flow/diffusion parameters. 
We have observed that using BP provides an explicit estimate not only for Z, but also for its first and second derivatives over the parameters. This observation has helped us to significantly accelerate the BP-based learning thus streamlining the search for the maximum of the partition
function over the parameters. The acceleration is achieved via the use of
Newton method in parallel with BP iterations. 
The trick itself gives an advantage to BP and it is currently 
not clear if similar accelerations are possible to achieve in the 
MCMC-based learning.

\section{Possible extensions}
\label{extensions}

In this section we discuss possible extensions of our approach that will 
be interesting to pursue in future work. We nevertheless find it useful 
to briefly discuss them here, in order to give a better idea about 
possibilities (and limitations) of a systematic probabilistic 
approach to particle tracking. 

\paragraph*{\bf Incorporating uncertainty}
Our approach allows simple modifications to incorporate various types
of uncertainty. For example, one can account for particles leaving and
entering the box (focal volume in the experimental realization) via
the following relaxation of the constraint $C\left(\{\sigma\}\right)$
in eq.~(\LI), \be C\left(\{\sigma\}\right)\equiv\prod_j\left[
  \delta\left(\sum_i\sigma_i^j,1\right) + e^{-\mu}
  \delta\left(\sum_i\sigma_i^j,0\right) \right] \prod_i\left[
  \delta\left(\sum_j\sigma_i^j,1\right) + e^{-\mu}
  \delta\left(\sum_j\sigma_i^j,0\right) \right]\, , \ee where
the parameter $\mu$ is interpreted as a chemical potential that accounts
for the fluctuations in the number of particles.  The BP equations
can be rewritten for this form of likelihood and 
the penalty parameter $\mu$ would have to be varied so as to optimize the 
likelihood and allow appropriate number of non-matching particles.

\paragraph*{\bf Processing of multiple snapshots}
Our approach can also be adopted to inference and learning based on
more than two subsequent snapshots. If the incremental displacement of
particles between two subsequent snapshots depends only on particle
positions at the early snapshot in the pair and does not depend on
other details of the preceding evolution of the particles, then
generalization of our approach consists in treating matchings between
each pair of snapshot in the sequence independently, stating the
total likelihood of the sequence as a product of the respective
expressions eq.~(\LI) for the pairs, and making global optimization 
over the governing parameters ${\bm\theta}$ for the entire product. 

If the motion of particles has memory spanning over $m$ snapshots, then our approach can still be implemented, however this will require knowledge or estimation for the joint probability of observations $P({\bm x_1},\dots,{\bm x_m}|{\bm \theta})$. Note that in this correlated case already the subtask of inferring MPA assignment, equivalent to the so-called multi-matching problem
\cite{MMR04,MartinMezard05}, becomes significantly more
difficult than in the non-correlated case.

Some sort of predictor-corrector scheme could also be investigated to solve the case with memory, or to achieve a speed-up of 
multiple snapshot processing. These promising possibilities are 
yet to be investigated and tested. 

\paragraph*{\bf Use of marginal probabilities}
We concentrated on inference of flow parameters in regime where tracking of individual particles is no longer possible. Note, however, that once the correct parameters are estimated the belief propagation provides also estimates of probabilities with which a given particle from the first snapshot moved to a given position in the second snapshot. This information may be useful in cases where we are interested in information about the trajectories of individual particles. The values of these probabilities also provide concrete information about the level of uncertainty where a given particle moved. 

Note, however, that if a particle is matched to the position of its
most probable image, the resulting configuration may even not be a
one-to-one mapping.  Indeed, a position in the second image might be
the most probable one for multiple particles from the first image.
Decimation-like techniques, of the type used to solve similar
inconsistency problems in the constraint satisfaction problems
\cite{MezardParisi02}, can then be used to obtain a consistent
one-to-one mapping.  In either case, even though the resulting
configuration may not be equal to the actual displacement, it may
still provide a useful visual information.

\paragraph*{\bf Learning multi-scale flow}
In this paragraph we describe how the learning framework, 
illustrated in the body of the manuscript on the model case 
of the Batchelor (single-scale) velocity, can be extended to the 
more general case of a multi-scale flow. The Batchelor model of velocity considered in the main text assumed that $\hat{s}$, the matrix of velocity gradients, does not depend on the particle index,  i.e. in other words, all the particles in the cloud sense the same velocity gradient.  In a more realistic turbulence setting the velocity gradient varies on the spatial scale correspondent to the viscous (Kolmogorov scale). Thus,  when particles are seeded in a cloud exceeding in size the viscous scale, one needs to use a richer higher-parametric model, for example stated in terms of the following harmonic expansion of the instantaneous velocity field: ${\bm u}({\bm r})=\sum_{\bm q}\exp(i{\bm q}{\bm r}) {\bm u}_{\bm q}$, where the number of harmonics (number of terms in the sum) is $N_q$. Then, velocity advecting particle $i$, where labeling is according to the first of the two consecutive images, is modeled in terms of the following gradient flow, ${\bm v}_i({\bm r}_i|{\bm r}^*_i)=\hat{s}_i({\bm x}_i)({\bm r}_i-{\bm r}_i^*)
+{\bm v}_i^{(0)}({\bm r}_i^*)$,
where ${\bm v}_i^{(0)}({\bm r}_i^*)={\bm u}({\bm r}_i^*)=\sum_{\bm q}\exp(i{\bm q}{\bm r}_i^*) {\bm u}_{\bm q}$, $\hat{s}_i^{(0);\alpha\beta}({\bm r}_i*)=
\nabla^\alpha u^\beta({\bm r}_i)=\sum_{\bm q}\exp(i{\bm q}{\bm r}_i^*) i q^\alpha
u^\beta_{\bm q}$, and ${\bm r}_i^*$ is a pre-selected and time-independent (frozen) grid point closest to the particle's initial position (in the first image), ${\bm x}_i={\bm r}_i(t=0)$. (One possible choice for ${\bm r}_i^*$ might be ${\bm x}_i$ itself.) The stochastic dynamic equation for trajectory of particle $i$ becomes
\begin{eqnarray}
\dot{\bm r}_i={\bm v}_i^{(0)}({\bm r}_i^*)+\hat{s}_i^{(0);\alpha\beta}({\bm r}_i^*)({\bm r}_i-{\bm r}_i^*)+{\bm \xi}_i(t).
\label{multi-scale_vel}
\end{eqnarray}
These equations can be integrated over time, thus resulting in the following generalization of Eqs.~(7-9) from the main text
\begin{eqnarray}
  &P_i^j({\bm x}_i,{\bm y}^j)=  ({\rm det}\, M_i)^{-\frac{1}{2}} \exp{\left( -\frac{1}{2} \tilde{\bm r}^\alpha(M^{-1})^{\alpha\beta}_i \tilde{\bm r}^\beta \right)}\,; \label{eq:prob1}\\
  &\tilde{\bm r}= {\bm y}^j-r_i^* - W_i(\int_0^\Delta dt'W^{-1}_i(t) {\bm v}_i^{(0)}+{\bm x}_i-{\bm r}_i^*); \\
  &M_i= \kappa\,  W_i(\Delta) \left[ \int_0^\Delta W^{-1}_i(t) W^{-1,T}_i(t) \, {\rm d}t \, \right] \,  W^T_i(\Delta)\, , \label{turb1}
\end{eqnarray}
where $W_i(t)=\exp(t\, \hat s_i)$. The task of reconstruction becomes to infer $N_q$ harmonics given positions of $N$ particles in the two snapshots,  
for example assuming that the diffusion coefficient $\kappa$ is known. Naturally, a reliable multi-parametric reconstruction is feasible if $N_q\ll N$. Extension of our BP-based scheme to this  multi-parametric setting is straightforward, as the graphical model employed to solve the problem is identical to the one described in the main text for the Batchelor model case. The main technical difficulty in learning the multi-parametric velocity will in fact be in maximizing the partition function of the model over a larger (than in the Batchelor case) number of the degrees of freedoms (harmonics). However, and as discussed above in Section \ref{BPMC}, this problem can be dealt with efficiently by alternating Newton (parameter adjustment) steps with BP iterative steps. The results and performance of the multi-scale inference are, however, yet to be tested. 

\paragraph*{\bf Interacting particles}
Let us note that extending our approach to systems of
interacting particles may be possible, but it would constitute a more
significant challenge, as in this case possible probabilistic models
of particles' evolution are much harder to evaluate.

\end{document}